# Safety Evaluation of Critical Applications Distributed on TDMA-Based Networks

François Simonot[1], Françoise Simonot-Lion[2], YeQiong Song[2]

[1] IECN – ESSTIN, UHP Nancy 1
Parc Robert Bentz – 54506 Vandœuvre-lès-Nancy (France)
[2] LORIA – INPL
Campus Scientifique, BP 239 – 54506 Vandœuvre-lès-Nancy (France)
simonot@esstin.uhp-nancy.fr, {simonot,song}@loria.fr

**Abstract.** Critical embedded systems have to provide a high level of dependability. In automotive domain, for example, TDMA protocols are largely recommended because of their deterministic behavior. Nevertheless, under the transient environmental perturbations, the loss of communication cycles may occur with a certain probability and, consequently, the system may fail. This paper analyzes the impact of the transient perturbations (especially due to Electromagnetic Interferences) on the dependability of systems distributed on TDMA-based networks. The dependability of such system is modeled as that of "consecutive-k-out-of-n:F" systems and we provide a efficient way for its evaluation.

## 1  Context

For achieving deterministic real time systems, the time-triggered approach is often used. In particular, in order to guarantee the real time properties on the exchanges between distant application components, one uses TDMA-based protocols. TDMA slots are assigned to each data producers in a periodic (or cyclic) way with a fixed TDMA cycle duration. A receiver node periodically receives therefore data it consumes at fixed TDMA time slots so that associated actions can be executed at the right time. Moreover an absence of data production or transfer can be easily detected by the system making the consumer to take a right decision if necessary. In practice, for providing more reliability, the designer of time-triggered applications generally makes data producers to send data to the consumer with period much smaller than necessary (i.e., Nyquist frequency), tolerating thus occasional production or transmission errors to some extend.  This is particularly interesting for systems operating at harsh conditions (i.e., subjected to environmental perturbations) provoking transient errors.

For instance, when TDMA-based networks are subjected to EMI (Electromagnetic Interference) perturbations, which are, for example, the typical case of automotive networks, message transmissions can be erroneous. In most of networks such as LIN,

TTP/C and FlexRay, transmission errors are detected but erroneous messages are not systematically retransmitted.

In this paper we focus on the analysis of the dependability of a TDMA-based network faced to EMI perturbations with respect to the dependability properties required by a safety critical application distributed on this network. In particular, examples of such an application are in-vehicle embedded systems. With the increasing deployment of electronic embedded components in vehicles, one hears in the media more and more vehicle failures due to the malfunctioning of electronic components.

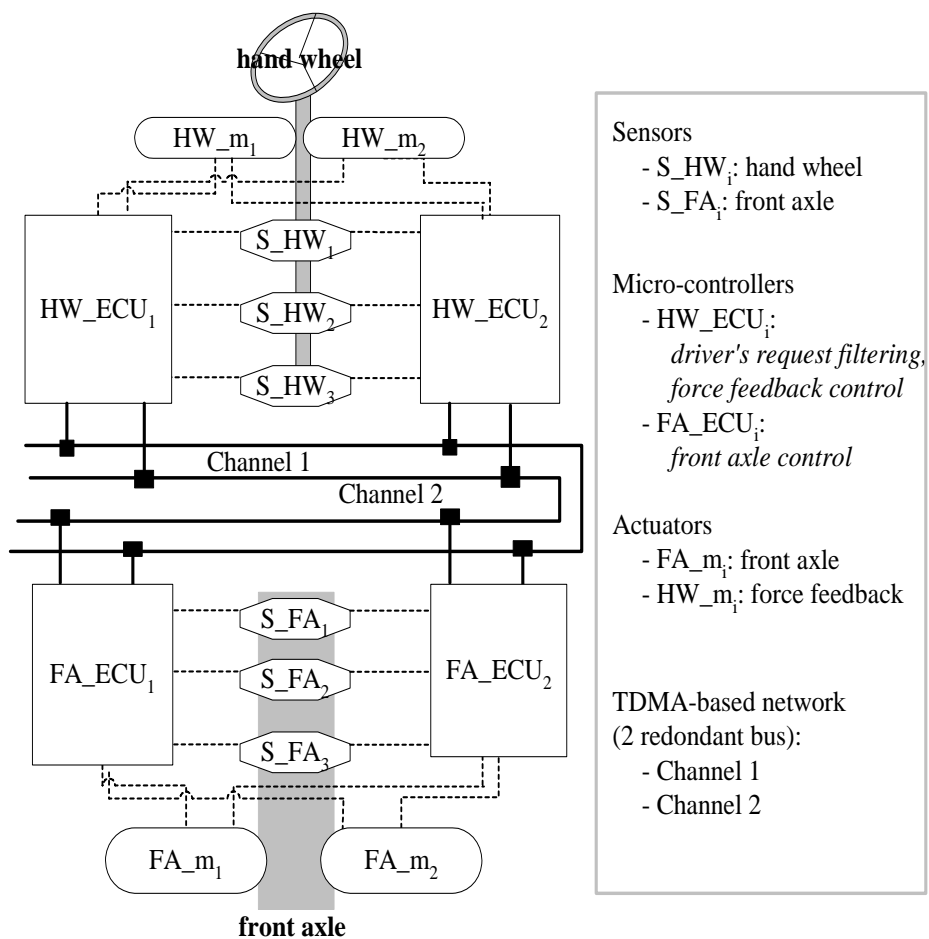

Fig. 1. A Steer-by-Wire architecture

If today's identified problems find rapidly their solutions, some non-identified problems, making random behaviors of the vehicle, remains unsolved (e.g. vehicle speed blocked by the cruise control system). It seems that EMI perturbations are the main cause of those random behaviors of the vehicle. The Wall Street Journal, edited 8 September 1997, quoted in Risk Digest, [9] talked about accidents involving vehicles, which to all appearances were caused by EMI. More recently in France, it happens that the cruise control system of certain cars behaves randomly (e.g. impossible to reduce the speed!). This problem can become even more critical with future integration of X-by-wire systems in a vehicle. So the behavior of such a system face to EMI has to be evaluated and in particular the impact of the network on the vehicle dependability has to be analyzed.

Let us consider a *Steer-by-Wire* system that aims to provide two main services: controlling the front axle actuation according to the driver's request and providing a "mechanical-like" force feedback to the hand wheel that is consistent with the current state of the vehicle. These services are assumed to be independent and we only focus on the "front axle actuation" service because it implies the most critical safety purpose. Fig. 1 represents the computer-based architecture. Because of its safety criticality, redundancy is omnipresent. Three redundant sensors, S_HW1, S_HW2, S_HW3, measure the driver's request (hand wheel angle and torque), two redundant actuators, FA_m1 and FA_m2, act on the front axle. Two redundant micro-controllers (ECU: Electronic Control Unit) are used for driver's requests filtering, HW_ECU1 and HW_ECU2 while two other redundant micro-controllers, FA_ECU1 and FA_ECU2, are dedicated to the support of the control laws for the front axle movement. Finally, three redundant sensors (S_FA1, S_FA2 and S_FA3) measure the state of the front axle and two redundant actuators, HW_m1 and HW_m2, provide the force feedback on the hand wheel. The four micro-controllers are connected on the redundant channels of a TDMA-like network (could be TTP/C or FlexRay). Critical functions such as front axle control are assumed to be executed periodically on FA_ECU1 and FA_ECU2, taking into account the output of the physical system to control, environmental information and the driver's request given at the hand wheel level. This last information is produced periodically by HW_ECU1 and HW_ECU2 and transmitted through a TDMA-based network to control law (Fig. 2). The length of the TDMA cycle is equal to the activation period of the control law ($T$ in Fig. 2).

We assume that the length of the TDMA cycle is less than the minimum interval between two driver's request transmissions. So the lack of driver's request samples, due for example to transmission errors, during a short-term, that is for a limited number of control law consecutive executions, can be tolerated. However, it is obvious that long-term absence of input data at the consumer side can lead to dangerous situations. So it is important to be able to evaluate the risk that *consecutive erroneous TDMA cycles* exceed the application-tolerating threshold (in terms of a given TDMA cycle number or equivalently the time length). This risk is termed in this paper the *application failure probability,* which gives a metric to measure the application robustness. Obviously this probability will depend on the perturbation model.

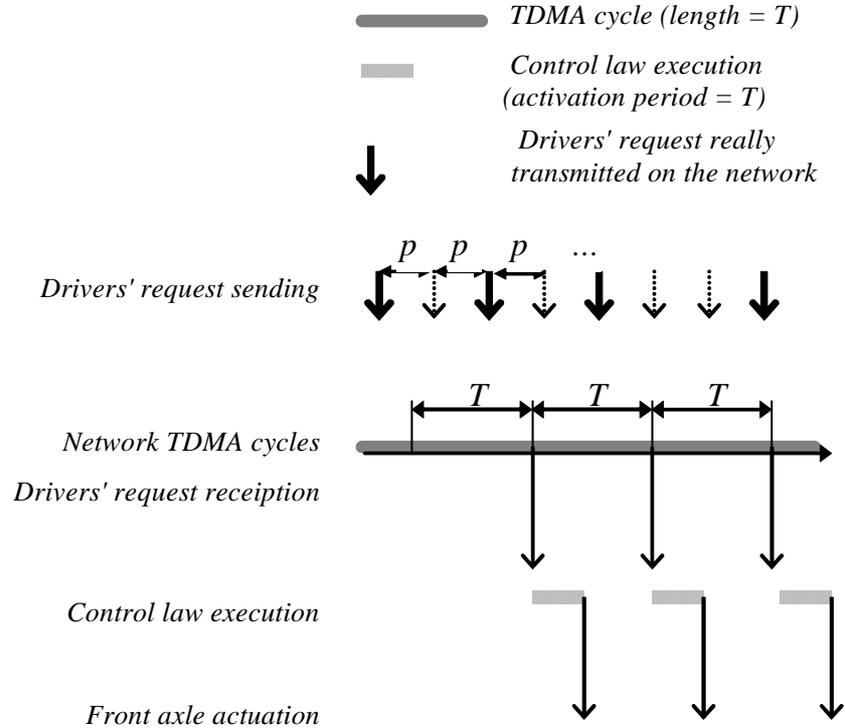

Fig. 2. Drivers' request transmission

In this paper, we propose a method for the evaluation of the *application failure probability* for critical systems distributed on a TDMA-based network and subject to several profiles of EMI perturbations. This evaluation is based on the assumption that one single non-erroneous cycle is enough for bringing the system to its normal state. This hypothesis is verified for the given Steer-by-wire architecture that we use as example in the following.

However, only little work has been done to deal with the *application failure probability* under EMI perturbations and more generally under transient faults. In [6] we first addressed the problem for applications distributed over CAN and introduced the worst-case deadline failure probability as one of the possible application dependability metrics. In [1], the impact of the EMI on the real-time delivery capability of CAN and TTCAN has been evaluated. These results cannot be readily applied to the TDMA-based networks. In fact, CAN retransmits whenever a transmission error occurs whereas in TDMA-based networks, there is no systematic retransmission upon transmission errors.

In our previous work (see [10] and [11]), we focused either on the evaluation of the X-by-wire application-tolerating threshold or the method to evaluate the application failure probability under restrictive hypothesis on the perturbation model. For instance, the probability for a TDMA cycle to be erroneous, termed in the following "TDMA cycle error probability", has been assumed constant. In fact it is not the gen-eral case. A given source can cause EMI that varies in time and / or the distance between a perturbation source and the vehicle can vary in time. Chao has demonstrated [3] that the evaluation of the application failure probability with constant TDMA cycle error probability can be performed based on the classic results on the "*consecutive-k-out-of-n:F*" systems. In this paper, we extend these results by dealing with the application failure probability evaluation with variable TDMA cycle error probability. So the previous theoretic results on the "*consecutive-k-out-of-n:F*" systems have to be extended.

In what follows, we will give in section 2 a description of the EMI perturbations that an in-vehicle application can meet; such perturbations may provoke TDMA cycle errors (because of either transmission errors or producing errors). Section 3 is devoted to our main contribution that is a method for evaluating the application failure probability with variable TDMA cycle error probability. Section 4 shows numerical applications for some typical error models (TDMA cycle error probability profiles) and that are obtained on a Steer-by-wire system. Section 5 gives the concluding remarks.

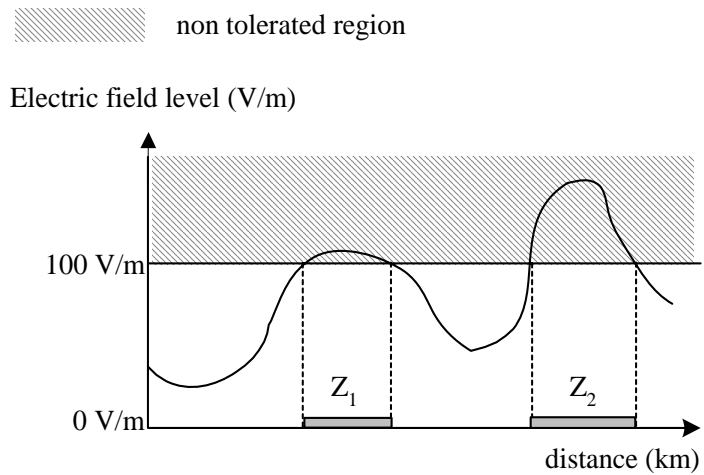

Fig. 3. Example of electric field level of a reference road

## 2   EMI perturbations and error models

Electromagnetic interferences are mainly caused by radio-communication transmitters, radars, and high voltage lines. Their influence on electronic compo-nents depends on the frequency and strength of the electromagnetic fields. In automotive industry, each carmaker specifies an internal regulatory policy that imposes the robustness level of electronic device with electromagnetic interference sources under a given voltage level and for a given interval of frequencies. So a test process is applied on each electronic component in order to verify its conformity to the specific carmaker standard.

Nevertheless, this conformity is just proved for given frequencies and voltage level. In fact, it is established that the testing condition are not met everywhere; it exists some traffic areas, for example near airports, where a vehicle can go through an area subject to a higher level of voltage and / or other frequencies than the specified ones and therefore, the probability that an in-vehicle embedded system can be corrupted by electromagnetic interferences is not zero. For example, carmakers often consider that the upper limit for the robustness assessment of electronic components is 100 V/m. This is to say that when a car goes through an EMI perturbed zone with a force higher than 100 volts per meter, its embedded electronic systems may exhibit errors.

Some sources of EMI are statically disposed along the road (for example, radars or high voltage lines). CEERF, a French project, funded by Ministry of Transport, proposed a characterization of the electromagnetic pollution for the French road system (Predit-CEERF, 2003). This project targeted mainly the automotive industry by proposing a cartography of the EMI sources and electromagnetic field levels in France and a method for its updating. These results are obtained thanks to a monitor embedded in a car and whose role is to record the frequency and the level of the ambient electromagnetic field during a journey along several representative roads. From this recording, we are able to select the length (in km) of each area under EMI perturbation of higher than 100 V/m (see Fig. 3); on the represented trajectory, two parts of this trajectory, areas $Z_1$ and $Z_2$, are subject to perturbations of more than 100V/m).

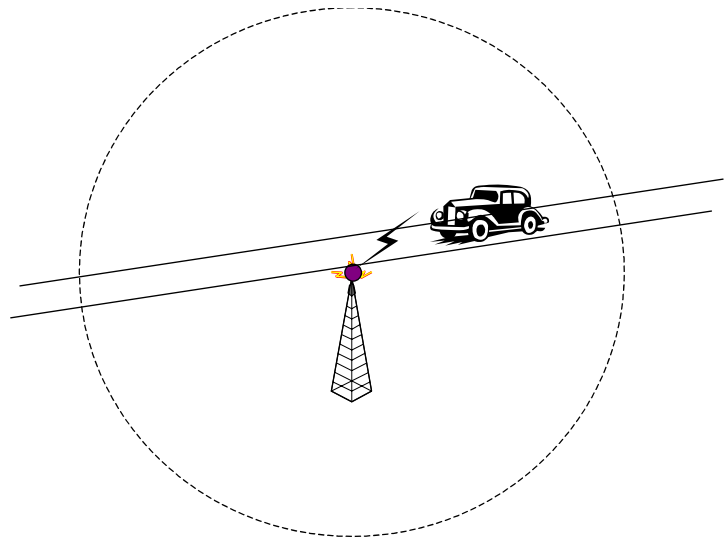

Fig. 4. Example of an EMI zone with variable

Without losing generality, in what follows, we will only focus on the analysis within *one* EMI zone. In fact, as assumed in section 1, for the targeted application, one sin-gle correct reception of the input data at the consumer side before the application-tolerating threshold is required for bringing the system to its normal state. So multiple zones (as they are independent each other) can be treated separately. In [10], we proposed how to evaluate the failure probability for a specific vehicle trajectory such as what is shown in Fig. 3.

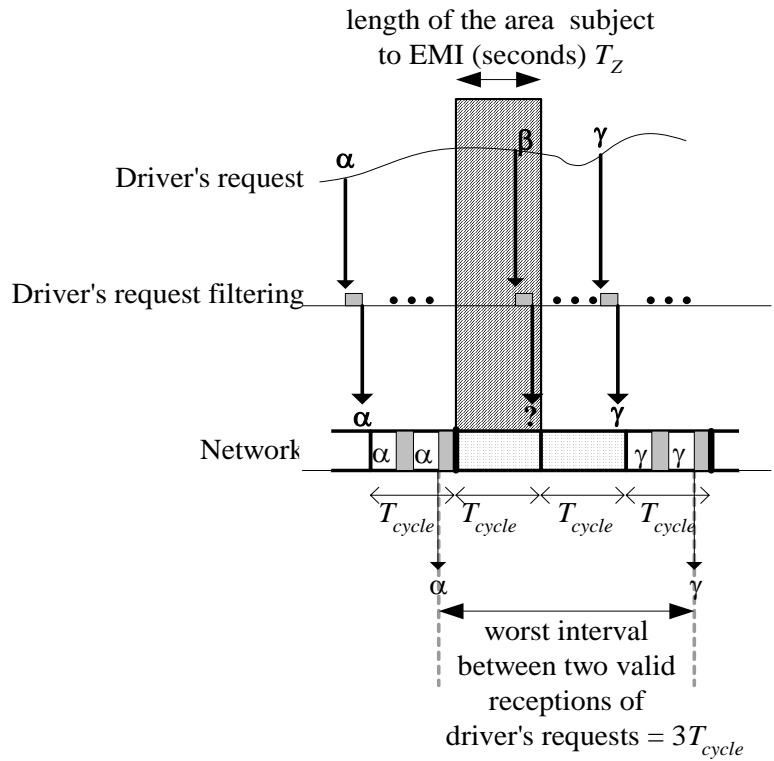

Fig. 5. Evaluation of the worst interval between two valid driver's requests

For passing through a given EMI area, a vehicle (assumed with a constant speed) will take a certain time called hereafter the *passing through time* and denoted by $T_z$. If one knows the TDMA cycle duration $T_{cyc}$, the passing through time can also be represented in terms of the number of TDMA cycles $n$. In [11] and [10], for taking into account the worst-case protocol recovery overhead, we evaluated this value. The first step consists in translating an area length expressed in meters in a length given in seconds (termed $Z$ in the following). Once this done, we have to evaluate how many cycles are possibly corrupted. Fig. 5 illustrates how to evaluate this in the worst case. When the filtering period is less than the TDMA cycle length (that is always the case of temporal redundancy through over-sampling of the hand wheel position), the worst case corresponds to the situation where all the replicated frames within the perturbation zone are corrupted and the end of the zone corrupts the beginning of the produc-tion of a TDMA cycle, causing thus an additional empty TDMA cycle. For the next valid TDMA cycle, as we assumed that the consumption takes place only after the last replica of a TDMA cycle, if this last replica is near the end of the TDMA cycle, it increases the worst interval between two valid driver's requests by still another additional cycle. So this worst-case interval is given by::

$$n = \left\lceil \frac{T_z}{T_{cyc}} \right\rceil + 2 \qquad (1)$$

If a rough approximation by the free space propagation model [8] can be used, the electric field strength at a point will be inversely proportional to the square of its distance to the source.

Unfortunately, the exact characterization of the TDMA cycle error probabilities within a zone has not been realized because of large measurements data needed for being statistically confident.

In this study, to get a general idea about the impact of EMI on the application robustness, we will evaluate the application failure probability by analyzing some typical profiles of $P = (p_1, p_2, ..., p_n)$ called hereafter "error models".

### 2.1 Constant-P model

This first error model describes a constant perturbation. For the total *passing through time* of $n$ TDMA cycles of a given EMI zone, we assume that each TDMA cycle has a same error probability, i.e. $p_i = p$ for all $i=1, 2, ..., n$.

Of course, this may not correspond to an actual situation. However, it is often the case of the laboratory tests. We keep it as a reference model for further comparisons.

### 2.2 Radio-P model

The second error model is a *Radio-P* model. It is designed to represent the error model of a vehicle passing through an EMI area of $n$ TDMA cycles generated by a radio transmitter (e.g. Fig. 3).

Assuming that the free space propagation model [8] is adopted and the error probability of a TDMA cycle is somehow proportional to the received electric field strength, for a given TMDA cycle $i$ ($i = 1, 2, ..., n$) we give its error probability by:

$$p_i = \frac{a}{(\frac{n+1}{2} - i)^2 + b} \qquad (2)$$

Where *a* and *b* (with $a \leq b$) are free parameters which can be adjusted for fitting to a concrete situation. This model is illustrated in Fig. 6.

This is only a general assumption. In practice, node and communication channel redundancy is often used to reduce the impact of EMI perturbations. The error probability estimation for a given perturbation could be more complex and field tests are necessary. In [10], we discussed in more detail the ways to estimate it.

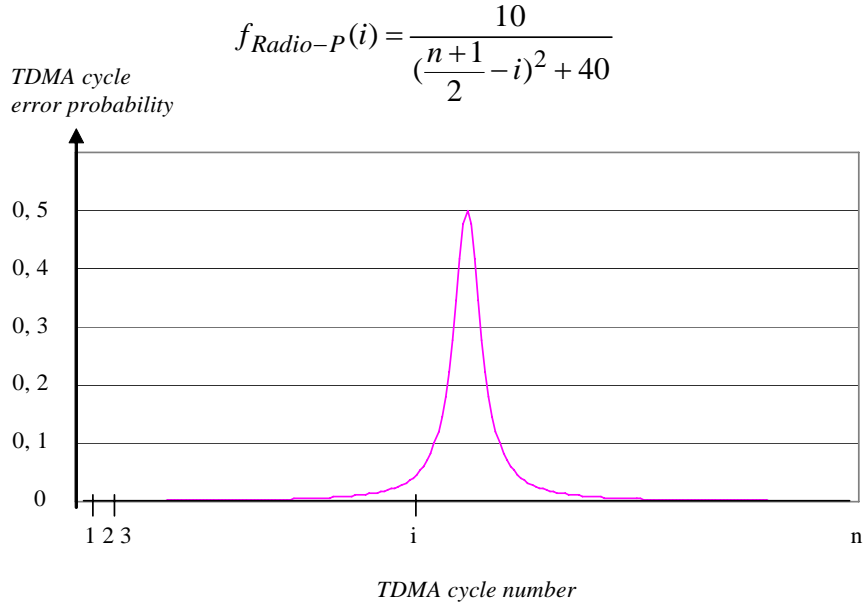

Fig. 6 Trends of *RadioP* model

### 2.3 Radar-P model

The third error model is a *Radar-P* model. It is proposed to represent the error model of a vehicle passing through an EMI zone of *n* TDMA cycles generated by a radar source (e.g. near to an airport).

The electric field varies periodically with the radar-scanning period of *T* (with $T < T_z$). Again we assume that the error probability of a TDMA cycle is proportional to the received electric field strength.

For a given TMDA cycle *i* (*i = 1, 2, ..., n*) we give its error probability by:

$$p_i = a + b \sin \frac{2\pi}{T} i \qquad (3)$$

Where *a* and *b* (with $a - b > 0$ and $a + b \leq 1$) are free parameters which can be adjusted for fitting to a concrete situation. Fig.4 depicts the trends of a *Radar-P* model.

$$f_{Radar-P}(i) = 0{,}1 + 0{,}09\sin\frac{2\pi}{375}$$

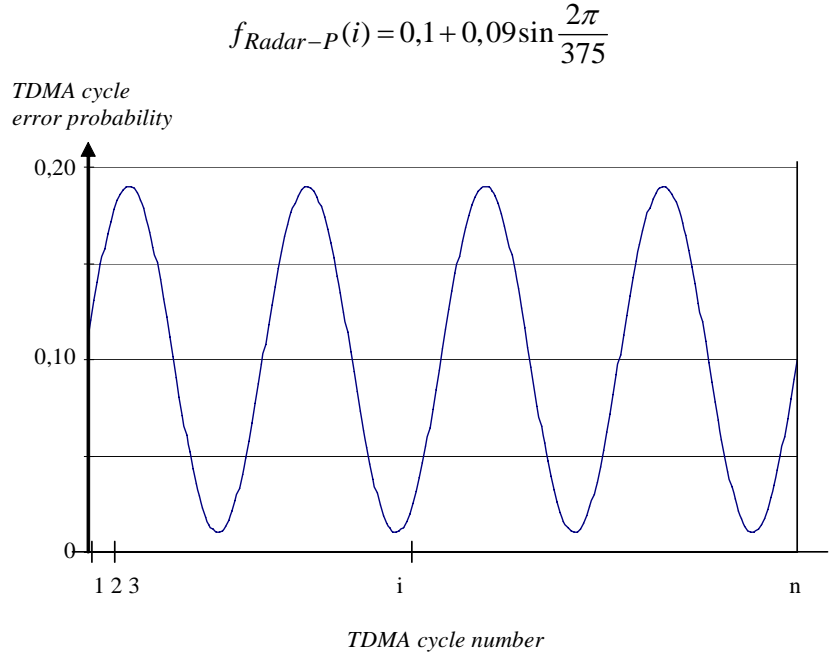

Fig. 7 Trends of *Radar-P* model

## 3 Application failure probability evaluation

### 3.1 Existing results for Constant-P model

Knowing a zone of *n* TDMA cycles and the application-tolerating threshold of *k* TDMA cycles, for constant *p*, the problem can be treated using existing results on the reliability of a system composed of an ordered sequence of *n* components and such that the system fails if and only if at least *k* consecutive components fail.

This kind of systems are termed "*consecutive-k-out-of-n:F*" systems and denoted by $C(k,n:F)$. For such a system, we note *n* the number of components, *p* the probability that a component fails, $L_n$ a number of consecutive failed components and *k-1* the largest tolerable number of consecutive failed components; the reliability of the system is evaluated by the probability that $L_n < k$, denoted by $P(L_n < k) = R(k,n;p)$, or equivalently the failure probability $P_{fail} = P(L_n \geq k) = 1 - R(k,n;p)$. The following formula was proposed first by Burr and Cane, in [2] and then simplified by Lambris and Papastavridis, in 1985 [5] and Hwang, in 1986 [4].

$$R(k,n;p) = \sum_{m=0}^{\lfloor (n+1)/(k+1) \rfloor} (-1)^m p^{mk} q^{m-1} \left( \binom{n-mk}{m-1} + q \binom{n-mk}{m} \right)$$

where $q = 1-p$.

The numerical evaluation of $R(k,n;p)$ via this formula is quite complex; so, we have developed in [10] a recurrent relation permitting to compute the failure probability for any $n$ and $k$.

For $P = (p_1, p_2, \ldots, p_n)$, as there does not exist closed form solutions, we give in the sequel a solution for $R(k,n;P)$.

### 3.2 New results for $P = (p_1, p_2, \ldots, p_n)$ with non constant $p_i$

Now this $P$, without loss of generality, let us consider an infinite sequence of independent Bernoulli trials $X_1, X_2, \ldots X_n, \ldots$ defined on the probability space $(\Omega, A, P)$ with $p_i = P(X_i = 1)$ for $i \geq 1$. We call "word" a sequence of consecutive successes of Bernoulli trials (when $X_i = 1$).

The goal of our work is to investigate the probability law of the random variable $L_n$, the length of the longest word known during the $n$ first trials. To our great surprise, it seems that this problem, in case of the non-identically distributed random variables, has never been addressed.

We define:

- $T_k$ = the first instant where a sequence of k consecutive successes appear.

- $u_n(k) = P(L_n < k)$ for $n \geq 0$ and $k \geq 1$. For a fixed value of $k$, $k \geq 1$, the sequence $u_n(k) = P(L_n < k)$ is decreasing and lower bounded by 0.

- $\lambda_n(k) = q_{n-k} p_{n-k+1} p_{n-k+2} \ldots p_n$ for $n \geq k$ with $q_0 = 1$ and $q_n = 1 - p_n$ if $n \geq 1$.

**Property1**

The sequence $u_n(k) = P(L_n < k)$ verifies the following relation:

For $k \geq 1$ and $n \geq k+1$,
$$u_n(k) = u_{n-1}(k) - \lambda_n(k) u_{n-k-1}(k) \tag{4}$$

with initial conditions :
$u_n(k) = 1$ for $0 \leq n \leq k-1$ and $u_k(k) = 1 - \lambda_k(k)$.

*Proof :*
We have :
$$P(T_k = n) = p_n p_{n-1} \ldots p_{n-k+1} q_{n-k} P(L_{n-k-1} < k)$$

and
$$P(L_n \geq k) = P(T_k \leq n),$$
giving the following relation :

$$u_n(k) = u_{n-1}(k) - \lambda_n(k) u_{n-k-1}(k) \text{ for } k \geq 1 \text{ and } n \geq k+1.$$
**End of proof**

*Property1* contains complete information on the behavior of $u_n(k) = P(L_n < k)$ and allows an exact calculation of the probability law of $L_n$ and provides an efficient algorithm for computing $u_n(k)$. Moreover, the following useful monotonic property of $u_n(k)$ can be established.

**Property2**

If $p'_j \geq p_j$ for all $j \geq 1$, then $u_n(k) \geq u'_n(k)$ for all $n$ and all $k \geq 1$.

That is to say : $p'_j \geq p_j$ for all $j \geq 1$ implies that $L_n$ is stochastically less or equal to $L'_n$ for all $n$. It turns out that $p'_j \geq p_j \geq p''_j$ for all $j \geq 1$ implies $u''_n(k) \geq u_n(k) \geq u'_n(k)$ for all $n$ and all $k \geq 1$.

**Algorithm**
The former recurrent relation can be implemented by the following algorithm:

```
// initialisation
for i=0 to k-1 do
   U(i)=1;
U(k)=1;
for i=1 to k
   U(k)=U(k)*p_i
U(k)=1-U(k)
// evaluation of P(L<k)
for j=k+1 to n
   λ=1-p_{j-k}
   for m=1 to k

   for m=1 to k
      λ=λ*p_{j-k+m}
```

```
    U(j)=U(j-1)-U(j-k-1)*λ
// P(L<k) is U(n)
```

## 4   Numerical results

In this section we will apply the previously established algorithms (the complexity of the program varies with n) to the three typical error models described above:
- Constant-P,
- Radio-P
- and Radar-P.

We focus on an EMI perturbation with passing through time $T_z=1500$ms and application-tolerating threshold $T_{max}=40$ms. Note that, for the Steer-by-wire system that we presented formerly, these values are extracted on the one hand, from the cartography of EMI sources and electromagnetic field levels in France obtained by the Predit project CERF in 2003 [7] and, on the other hand, by executing a Matlab / Simulink model that integrates the control law, the physical system and the vehicle characteristics [10].

For a given TDMA cycle duration $T_{cyc}$, n is given by equation 1, whereas the application-tolerating threshold in terms of the number of TMDA cycles, k, is given by:

$$k = \left\lfloor \frac{T_{max}}{T_{cyc}} \right\rfloor \quad (5)$$

Our objective is to evaluate the application failure probability $P_{fail}$ for a given TDMA cycle duration $T_{cyc}$.

In order to also analyze the influence of the TDMA cycle duration on $P_{fail}$, we make vary $T_{cyc}$ and, consequently, the activation period of the control law from 4ms to 10ms with step of 0.25ms (these values have to be specified both by automatic control specialists and by system architect designer). The obtained results provide guidelines for system designer to correctly dimensioning $T_{cyc}$ for meeting a specific requirement on $P_{fail}$.

### 4.1   Constant-P model

Let $p = 0.1$, by using the algorithm given by property1, we get the following failure probability $P_{fail}$ (Table 1).

In view of the equations 1 and 5, when $T_{cyc}$ increases, n and k both decrease. As was seen in section 3.2, for a fixed k value, the failure probability is an increasing function of n.

**Table 1** Application failure probability under *Constant-P* error model

| TDMA Cycle Length $T_{cyc}$ | Application failure $P_{fail}$ | Number of TDMA Cycles n | Maximum tolerable number of consecutive erroneous cycles k |
|---|---|---|---|
| 4 | 3.30E-09 | 377 | 10 |
| 4.25 | 3.12E-08 | 355 | 9 |
| 4.5 | 2.95E-07 | 336 | 8 |
| 4.75 | 2.79E-07 | 318 | 8 |
| 5 | 2.65E-07 | 302 | 8 |
| 5.25 | 2.53E-06 | 288 | 7 |
| 5.5 | 2.41E-06 | 275 | 7 |
| 5.75 | 2.31E-05 | 263 | 6 |
| 6 | 2.21E-05 | 252 | 6 |
| 6.25 | 2.12E-05 | 242 | 6 |
| 6.5 | 2.04E-05 | 233 | 6 |
| 6.75 | 1.98E-04 | 225 | 5 |
| 7 | 1.91E-04 | 217 | 5 |
| 7.25 | 1.84E-04 | 209 | 5 |
| 7.5 | 1.77E-04 | 202 | 5 |
| 7.75 | 1.72E-04 | 196 | 5 |
| 8 | 1.67E-04 | 190 | 5 |
| 8.25 | 0.00161977 | 184 | 4 |
| 8.5 | 0.00157484 | 179 | 4 |
| 8.75 | 0.0015299 | 174 | 4 |
| 9 | 0.00148497 | 169 | 4 |
| 9.25 | 0.00144902 | 165 | 4 |
| 9.5 | 0.00140408 | 160 | 4 |
| 9.75 | 0.00136813 | 156 | 4 |
| 10 | 0.00133218 | 152 | 4 |

## 4.2 Radio-P model

Let $a = 10$ and $11$, $b = 20$ and $19$ respectively, $p_i$ are given according to equation 2. The failure probability $P_{fail}$ for different $T_{cyc}$ is given in Table 2. In addition to the general comments we have already made for the *Constant-P* case, we can also observe the effect of property 2.

**Table 2** Application failure probability under *Radio-P* error model

| TDMA-Cycle Length $T_{cyc}$ | Application failure (a=10, b=20) $P_{fail}$ | Application failure (a=11, b=19) $P'_{fail}$ | Number of TDMA Cycles n | Maximum tolerable number of consecutive erroneous cycles k |
|---|---|---|---|---|
| 4 | 2.22E-08 | 8.19E-08 | 377 | 10 |
| 4.25 | 2.94E-07 | 9.73E-07 | 355 | 9 |
| 4.5 | 3.30E-06 | 9.82E-06 | 336 | 8 |
| 4.75 | 3.30E-06 | 9.82E-06 | 318 | 8 |
| 5 | 3.30E-06 | 9.82E-06 | 302 | 8 |
| 5.25 | 3.12E-05 | 8.32E-05 | 288 | 7 |
| 5.5 | 3.12E-05 | 8.32E-05 | 275 | 7 |
| 5.75 | 2.46E-04 | 5.86E-04 | 263 | 6 |
| 6 | 2.46E-04 | 5.86E-04 | 252 | 6 |
| 6.25 | 2.46E-04 | 5.86E-04 | 242 | 6 |
| 6.5 | 2.46E-04 | 5.86E-04 | 233 | 6 |
| 6.75 | 0.001609891 | 0.00340238 | 225 | 5 |
| 7 | 0.001609891 | 0.00340238 | 217 | 5 |
| 7.25 | 0.001609891 | 0.00340238 | 209 | 5 |
| 7.5 | 0.001609891 | 0.00340238 | 202 | 5 |
| 7.75 | 0.001609891 | 0.00340238 | 196 | 5 |
| 8 | 0.001609891 | 0.00340238 | 190 | 5 |
| 8.25 | 0.008690406 | 0.01621666 | 184 | 4 |
| 8.5 | 0.008690406 | 0.01621666 | 179 | 4 |
| 8.75 | 0.008690406 | 0.01621666 | 174 | 4 |
| 9 | 0.008690406 | 0.01621666 | 169 | 4 |
| 9.25 | 0.008690406 | 0.01621666 | 165 | 4 |
| 9.5 | 0.008690406 | 0.01621666 | 160 | 4 |
| 9.75 | 0.008690406 | 0.01621666 | 156 | 4 |
| 10 | 0.008690406 | 0.01621666 | 152 | 4 |

In fact, if we note:   $P = (p_1, p_2, ..., p_n)$ for $a = 10$ and $b = 20$,
$P' = (p'_1, p'_2, ..., p'_n)$ for $a = 11$ and $b = 19$,
According to equation 2, it turns out that $p'_i > p_i$ for all $i = 1, 2, ..., n$. So it is not surprising that $P'_{fail} > P_{fail}$ in Table 2.

**Table 3** Application failure probability under *Radar-P* error model

| TDMA-Cycle Length $T_{cyc}$ | Application failure $P_{fail}$ | Number of TDMA Cycles n | Maximum tolerable number of consecutive erroneous cycles k |
|---|---|---|---|
| 4 | 5.55E-07 | 377 | 10 |
| 4.25 | 2.93E-06 | 355 | 9 |
| 4.5 | 1.57E-05 | 336 | 8 |
| 4.75 | 1.47E-05 | 318 | 8 |
| 5 | 1.38E-05 | 302 | 8 |
| 5.25 | 7.53E-05 | 288 | 7 |
| 5.5 | 7.14E-05 | 275 | 7 |
| 5.75 | 3.92E-04 | 263 | 6 |
| 6 | 3.74E-04 | 252 | 6 |
| 6.25 | 3.57E-04 | 242 | 6 |
| 6.5 | 3.42E-04 | 233 | 6 |
| 6.75 | 0.00192067 | 225 | 5 |
| 7 | 0.00184522 | 217 | 5 |
| 7.25 | 0.0017695 | 209 | 5 |
| 7.5 | 0.00170302 | 202 | 5 |
| 7.75 | 0.00164584 | 196 | 5 |
| 8 | 0.00158847 | 190 | 5 |
| 8.25 | 0.00907411 | 184 | 4 |
| 8.5 | 0.008806 | 179 | 4 |
| 8.75 | 0.00853722 | 174 | 4 |
| 9 | 0.00826772 | 169 | 4 |
| 9.25 | 0.00805156 | 165 | 4 |
| 9.5 | 0.0077806 | 160 | 4 |
| 9.75 | 0.0075632 | 156 | 4 |
| 10 | 0.00734519 | 152 | 4 |

This kind of numerical results can be used to verify whether a given application, distributed on TDMA-based networks and under a known EMI zone, can still meet the dependability constraint in terms of failure probability. It can also provide to a designer with guidelines for correctly dimensioning $T_{cyc}$ for meeting a specific requirement on application failure probability.

## 5 Conclusion

In this paper we have investigated the impact of the EMI perturbations on the dependability of applications distributed around TDMA-base networks where we assumed that application failure occurs when consecutive erroneous TDMA cycles exceed a certain threshold. This problem is of prime importance, especially for automotive industry as on the one hand, the most adopted embedded networks such as LIN, CAN and FlexRay are based on TDMA, and on the other hand many embedded applications (e.g. X-by-Wire systems) have to meet stringent dependability constraints, this even under EMI perturbations. We contributed to the method for evaluating the application failure probability. For this, we have proposed an important theoretic result which extends the existing one on the "*consecutive-k-out-of-n:F*" systems to including variable probability. Although we have only analyzed several typical error models, our method is still available whatever the profile of $p_i$ may be. These probabilities can be obtained in practice by measurements. This method can also be used to study the system dependability of any transient perturbations.